\documentclass[a4paper, prb, twocolumn, showpacs, superscriptaddress]{revtex4}

\usepackage{amssymb}
\usepackage{amsmath}
\usepackage{color}
\usepackage{graphics, graphicx}
\usepackage{bbold}
\usepackage{psfrag}
\usepackage{mathcomp}

\newcommand{\IM}{\mbox{Im}}

\begin{document}

\author{D. Semmler} 
\affiliation{Institut f\"ur Theoretische Physik, Johann Wolfgang Goethe-Universit\"at, 60438 Frankfurt am Main, Germany}
\author{K. Byczuk}
\affiliation{Institute of Theoretical Physics, Warsaw University, ul. Ho\.{z}a 69, 00-681 Warszawa, Poland}
\affiliation{Theoretical Physics III, Center for Electronic Correlations and Magnetism, Institute for Physics, University of Augsburg, 86135 Augsburg, Germany}
\author{W. Hofstetter}
\affiliation{Institut f\"ur Theoretische Physik, Johann Wolfgang Goethe-Universit\"at, 60438 Frankfurt am Main, Germany}

\date{\today}

\title{Mott-Hubbard and Anderson metal-insulator transitions in correlated lattice fermions with binary disorder}

\begin{abstract}
Strongly correlated fermions in a crystal or in an optical lattice in the presence of binary alloy disorder are investigated. We employ the statistical dynamical mean-field theory, which incorporates both, fluctuations due to disorder and local correlations due to interaction, to solve the Anderson-Hubbard model. Localization due to disorder is studied by means of the probability distribution function of the local density of states. We obtain a complete paramagnetic ground state phase diagram consisting of disordered correlated metal, Anderson-Mott insulator, and band insulator. 
\end{abstract}

\maketitle

\section{Introduction}

The interplay between disorder and interaction in correlated electron materials still remains far from complete understanding \cite{Mott90} in spite of major progress achieved in disordered but non-interacting electron systems.\cite{Anderson58, Kramer93} On the theoretical side, the main obstacle is the non-perturbative character of the most interesting phenomena, e.g., alloy-band splitting, Anderson localization, or the Mott-Hubbard metal-insulator transition.\cite{Imada98} On the experimental side, the main problem is that both the amount of disorder and the strength of the interaction are not well controlled and not easily tunable in real condensed matter systems. The idea of a quantum simulator,\cite{Feynman82} where a complicated quantum many-body system is simulated by another quantum but perhaps simpler system working as a quantum computer, is very attractive. Regarding the interplay between disorder and interaction, experiments with ultracold fermionic or bosonic atoms in optical lattices\cite{Jaksch98,Greiner02,Hofstetter02,Koehl05,Joerdens08,Schneider08} are very promising steps toward creating such a quantum simulator and, therefore, are capable of shedding light into this unsettled problem. 

Disorder in ultracold gases can be simulated in different ways: (i) by using an optical speckle laser,\cite{Billy08,DeMarco08} (ii) by superimposing two laser beams with incommensurate frequencies,\cite{Fallani07} or (iii) by loading two atomic species, where only one is mobile, into an optical lattice.\cite{Guenter06,Ospelkaus06} The latter simulates a binary-alloy distribution of the on-site energies. Effects of interactions in optical lattices are controlled by tuning the on-site potential depths and/or the magnetic field around a Fesh\-bach resonance.\cite{Theis04} Recently developed momentum resolved radio frequency (rf) spectroscopy,\cite{Stewart08} on the other hand, is an adequate probing technique of correlated and disordered systems. This technique is similar to the well-known angle resolved photoemission spectroscopy experiments,\cite{Chen09,arpes} which were performed for alloys in solid-state physics. rf spectroscopy probes the spectral function and, thereby, the single-particle Green's function of the many-body system. 

Strongly correlated fermions in three dimensions are successfully described within the dynamical mean-field theory (DMFT).\cite{Metzner89,Vollhardt93,Georges96} This mean-field theory is fully non-perturbative and in combination with density functional theory it is capable of describing properties of real solid-state systems.\cite{Kotliar06} An extension of DMFT, which includes disorder effects, was performed both, in analogy to the well-known coherent potential approximation (CPA),\cite{Vlaming92,Ulmke95} and within a fully stochastic approach to incorporate effects of Anderson localization.\cite{Dobrosavljevic97,Song08} Since the latter approach is  computationally very expensive if one treats correlation effects on a rigorous level and keeps sufficiently large ensembles of disorder realizations, the typical medium theory (TMT-) DMFT was developed.\cite{Dobros2003} Here the geometrically averaged local density of states (LDOS) is used as an order parameter for Anderson localization. TMT-DMFT was successfully applied to the non-interacting\cite{Dobros2003} and to the 
interacting\cite{Byczuk05,Byczuk05_a,Dobro2009,Byczuk09} electron systems with disorder.

By construction, TMT-DMFT is only capable of describing effects of strong localization due to disorder, i.e., effects caused by fluctuations of the wave-function amplitudes. Since TMT-DMFT determines the typical LDOS, i.e. the most probable value of the LDOS, all non-local phase interference effects are missed. To improve the theory such that weak-localization effects in the many-particle wave function are kept, one should combine a fully stochastic approach with DMFT.\cite{Dobrosavljevic97} In this way one can readopt the original point of view of Anderson \cite{Anderson58} and use the full probability distribution function (PDF) of the LDOS as an order parameter for the Anderson transition within DMFT. Recently, also the periodic Anderson model with disorder was investigated by means of the statistical DMFT.\cite{Miranda01a,Miranda01b,Aguiar03} Therein, a novel electronic Griffith's phase, characterized by non-Fermi-liq
uid behavior, was established as a precursor of a disorder-driven metal-insulator transition. For non-interacting disordered systems such a stochastic theory, named local distribution (LD) approach, was effectively used in an analytical approach\cite{Abou73} and recently implemented numerically.\cite{Alvermann05}

The aims of this paper are to apply the statistical DMFT\cite{Dobrosavljevic97,Alvermann05} to interacting and disordered fermions and to extend the method to a level not reached so far. The theory is applied to correlated fermions on a lattice with binary-alloy type of disorder. This problem has recently been addressed within DMFT combined with CPA to deal with disorder.\cite{Byczuk04} It was shown, in particular, that new types of alloy-Mott or alloy-charge transfer insulators can appear and that the Mott-Hubbard metal-insulator transition can occur at non-integer particle densities.\cite{Byczuk04} Here we revisit this model and show that Anderson localization significantly extends the picture. In order to make the statistical DMFT method computationally feasible the DMFT part is solved approximately within modified perturbation theory (MPT),\cite{Kajueter97, Potthof97} which (in contrast to, e.g. the slave boson mean-field theory\cite{Kotliar86,Georges96}) provides a reliable interpolation scheme between the weakly and strongly interacting regimes.\cite{Zhang93} However, a quantitative analysis of the above mentioned Griffiths phase is limited by MPT, as this impurity solver does not reproduce the exponentially small low-energy scale for strong interactions.  

The paper is structured as follows: In Sec. \ref{system} we introduce and motivate the underlying physical model. The LD method for non-interacting disordered systems is reviewed in Sec. \ref{no-ia} and extended to the statistical DMFT for interacting systems in Sec. \ref{ia}. Our main results are discussed in Sec. \ref{results}. Finally, the connection to experiments in optical lattices is discussed in Sec. \ref{connection_Exp}.

\section{Anderson-Hubbard model with binary-alloy disorder}\label{system}
Electrons or cold fermionic atoms, such as $^6$Li or $^{40}$K, in disordered lattices are well described by the Anderson-Hubbard Hamiltonian 
\begin{equation}
 H = -\sum\limits _{ i j \sigma } t_{ij} c_{i \sigma}^{\dagger} c_{j \sigma}   
 - \sum \limits_{i \sigma} (\mu - \epsilon_i) c_{i \sigma}^{\dagger} c_{i \sigma} + U \sum \limits_{i} n_{i \uparrow } n_{i \downarrow }  \label{eq_sys1}
\end{equation}
where $c_{i \sigma}^{\dagger}$  ($c_{i \sigma}$) denotes creation (annihilation) operators at a lattice site $i$ with spin $\sigma=\pm 1/2$. The fermionic number operator is given by $n_{i \sigma}= c_{i \sigma}^{\dagger}c_{i \sigma}$. The hopping amplitude between sites $i$ and $j$ is denoted by $t_{ij}$, the interaction amplitude is represented by $U$, and the chemical potential is given by $\mu$. In the following we consider fermions on a Bethe lattice\cite{Georges96} with connectivity $K$, which is related to the coordination number $Z$ via $K=Z-1$, where the hopping amplitude is only nonzero  $t_{ij}=t$ for nearest neighbors $i$ and $j$. We also set energy units such that the band-width $W_0=4t\sqrt{K}=1$ hereafter. The local disorder is given by random on-site energies $\epsilon_i$, which are drawn from a probability distribution function $p_{\epsilon}(\epsilon _i)$.

In this paper we consider the case of a binary-alloy Anderson-Hubbard model, in which the PDF of the on-site energies is given by the bimodal function 
\begin{equation}
p_{\epsilon}(\epsilon _i) = x \delta(\epsilon _i+\frac{\Delta}{2}) + (1-x) \delta(\epsilon _i-\frac{\Delta}{2}),  \label{sysPD}
 \end{equation}
where $x$ and $1-x$ are the fractions of lattice sites with energies $\epsilon _i=-\frac{\Delta}{2}$ and $ \epsilon _i=\frac{\Delta}{2}$, respectively, and $\Delta$ describes the on-site energy splitting. In general, $\Delta$ and $x$ are independent parameters. However, the cases $x=0$ or $1$ correspond to non-disordered systems with on-site energy shift $\pm \Delta/2$. Therefore, a natural parameter for measuring the disorder strength in binary alloy systems is $\delta\equiv x(1-x)\Delta$.\cite{Yu08}

A very important difference between binary-alloy disorder and disorder types with continuous probability distributions is that in the former case in a non-interacting system and in arbitrary lattices the Bloch band is split if $\Delta>W_0$.\cite{Kirkpatrik70,Gonis,Byczuk04} In this limit two alloy subbands are formed and the system is a band insulator if $\nu=2x$ or $\nu=2$, where $\nu$ is number of fermions per site, or a metal otherwise. In the presence of interaction a Mott insulator at fractional particle filling $\nu=x$ or $\nu=1+x$ is allowed.\cite{Byczuk04,Byczuk03} Here we investigate how Anderson localization modifies these predictions. 

\begin{figure}[tb]
\includegraphics[width=7cm]{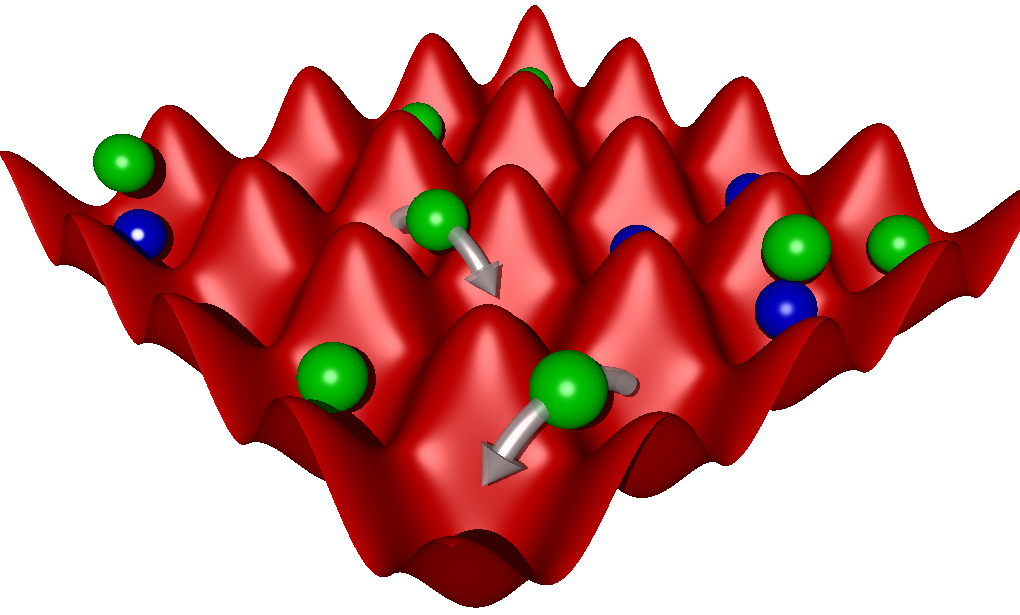}  
\caption{
(Color online) Illustration of a realization of binary disorder in optical lattices. Two atomic species (indicated as light green and dark blue spheres) are loaded into an optical lattice. The hopping amplitude of one species (dark blue) is suppressed and therefore these atoms are immobile. Due to the interatomic interaction the second species experiences a binary disordered lattice potential depending on the presence of an atom of the immobile species on the same lattice site, i.e. if there is a dark blue atom present the on-site energy is $\epsilon_i=+\Delta / 2$ otherwise $\epsilon_i=-\Delta / 2$.}
\label{figLattice}
\end{figure}

In systems of cold atoms in optical lattices the binary-alloy disorder is prepared by adding an additional species of atoms, which are immobile but  interact with the mobile components. First experimental attempts in this direction have been performed.\cite{Guenter06,Ospelkaus06} However, in such a system one must take care that the immobile atom positions are random but not fluctuating in time, i.e. the created disorder must be quenched.\cite{comment} Such a situation is schematically presented in Fig.~\ref{figLattice}.

\section{Method}\label{method}

In this section we introduce the notation used and describe the statistical DMFT for non-interacting and for interacting systems in Secs. \ref{no-ia} and \ref{ia}, respectively.

\subsection{Local distribution approach}\label{no-ia}

The local distribution approach is a self-consistent computational scheme for determining the probability distribution function of the local single-particle Green's functions, i.e. $p\left[G_{ii\sigma}(\omega)\right]$. Here $G_{ij\sigma}(\omega)$ is the Fourier transformation of the retarded Green's function $G_{ij\sigma}(t)=-i\theta(t)\langle [ c_{i\sigma}(t),c_{j\sigma}^{\dagger}(0) ]_+ \rangle$, where $\theta(t)$ is a Heaviside function and $[..,..]_+$ denotes anticommutator brackets. In the following we consider only paramagnetic solutions of the Anderson-Hubbard model and therefore the spin index $\sigma$ is omitted.

In the absence of interactions the renormalized perturbation theory\cite{Economou_book} shows that the local Green's function can always be expressed as
\begin{equation}
G_{ii} (\omega) = \frac{1}{\omega + \mu -\epsilon_i - \Gamma_i(\omega) + i \eta} \,, \label{eq_M1}
\end{equation} 
where the hybridization function $\Gamma_i(\omega)$ describes all effects of the coupling of site $i$ with other nearest neighbor lattice sites. The chemical potential is given by $\mu$.  For numerical reasons we also introduced the broadening factor $\eta>0$. In order to study localization effects, the limit $\eta\rightarrow0$ has to be performed. 

The hybridization function $\Gamma_i(\omega)$ can be expressed by an infinite, renormalized series of the form
\begin{equation}
\Gamma_i (\omega) = t^2 \sum\limits_{j=1}^{K} G_{ji}^{(i)}(\omega)G_{ij}^{(i)}(\omega) + \cdot\cdot\cdot ,\label{eq_M2}
\end{equation} 
where $G_{ji}^{(i)}(\omega)$ is the cavity Green's function of the system when the site $i$ is removed. 

On the Caley tree (Bethe lattice) this series can be exactly truncated after the first term and the hybridization function is exactly given by\cite{Eckstein,Kollar}
\begin{equation}
\Gamma_i (\omega) = t^2 \sum\limits_{j=1}^{K} G_{jj}^{(i)}(\omega),\label{eq_M3}
\end{equation} 
with $K$ as the coordination number of the lattice.
Here we use the fact, that the geometry of the Bethe lattice does not change when site $i$ is removed, which allows us to determine the cavity Green's function $G_{jj}^{(i)}$ in analogy to $G_{ii}$.

\begin{figure}[tb]
\includegraphics[width=4.5cm]{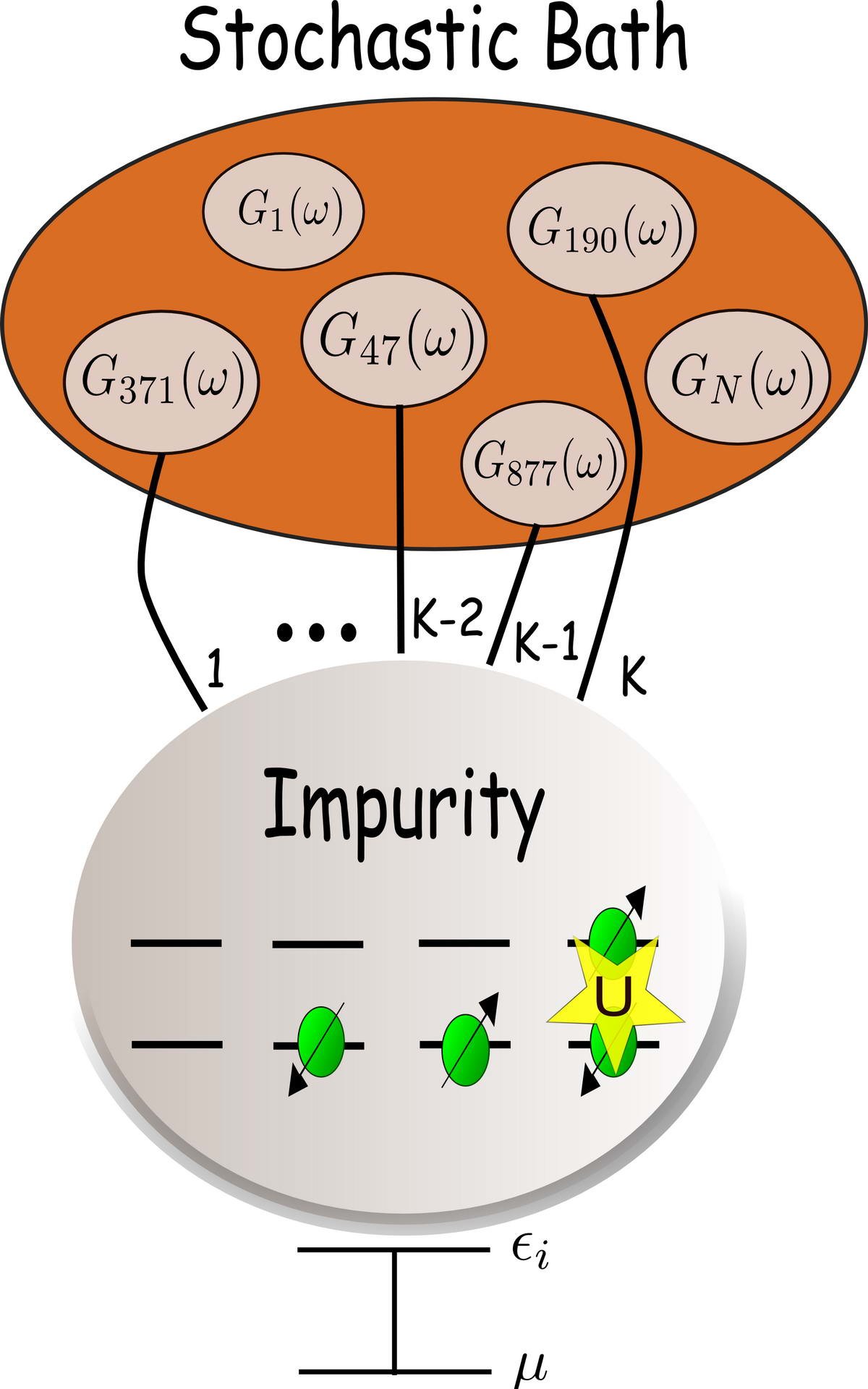}  
\caption{(Color online) Illustration of the statistical dynamical mean-field theory applied in this work. The many-body problem with disorder is mapped onto an ensemble of single impurities, which are coupled to an ensemble of stochastic Green's functions, which is determined self-consistently. $G_n$ represents the $n$th sample from the ensemble of Green's functions. 
}
\label{figMethod}
\end{figure}

In practice, given an initial PDF $p\left[G_{ii}(\omega)\right]$  the computational scheme is the following: (i) For each ensemble member we draw a random on-site energy $\epsilon_i$ out of the PDF $p_{\epsilon}(\epsilon_i)$ given in Eq.~(\ref{sysPD}). (ii) The hybridization function $\Gamma_i(\omega)$ is determined via Eq.~(\ref{eq_M3}), in which the nearest neighbor cavity Green's functions $G_{jj}^{(i)}(\omega)$ are randomly sampled from the PDF $p\left[ G_{ii}(\omega)\right]$. (iii) The local single-particle Green's function $G_{ii}(\omega)$ is calculated using Eq.~(\ref{eq_M1}). (iv) Having calculated all new $G_{ii}(\omega)$ a new PDF $p\left[G_{ii}(\omega)\right]$ is obtained and we return to step (i). The algorithm is repeated until self-consistency for $p\left[ G_{ii}(\omega)\right]$ is achieved. We note that this method incorporates spatial fluctuations, i.e. quantum interference effects, caused by the disorder. Schematically the computational procedure is presented in Fig.~\ref{figMethod}.

The relevant physical observable is the LDOS $\rho_i (\omega) = - \frac{1}{\pi} \IM(G_{ii}(\omega))$, which is a random quantity in disordered systems. The corresponding distribution $p[\rho_i (\omega)]$ is obtained by counting all values of the LDOS for each frequency and constructing a  histogram.\cite{comment_stat} 
From this probability distribution we can then determine the expectation value, i.e. the arithmetically averaged LDOS
\begin{equation}
\langle \rho (\omega) \rangle_{\mbox{\tiny arith}} = \langle \rho_i (\omega) \rangle_{\mbox{\tiny dis}} 
\end{equation}
and the typical value, which we approximate by the geometrical average
\begin{equation}
\langle \rho (\omega) \rangle_{\mbox{\tiny geom}} =  \exp \langle \ln \rho_i (\omega) \rangle_{\mbox{\tiny dis}},
\end{equation}
where $\langle F \rangle_{\mbox{\tiny dis}} = \int_0^{\infty} d x F(x) p[x]$ is the average over different disorder realizations of the corresponding quantity $F$.
In the following, the cumulative probability distributions 
\begin{equation}
P[\rho (\omega)] =  \int\limits_0^{\rho(\omega)} p[\rho\prime(\omega)] d\rho\prime (\omega) \,
\end{equation}
will also be useful to characterize the disordered system.

\subsection{Interacting systems}\label{ia}

In the presence of interactions, Eq.~(\ref{eq_M1}) is no longer true. It should be replaced by a Dyson-like equation, which relates the inverse of $G_{ij}(\omega)^{-1}$ with the self-energy functions $\Sigma_{ij}(\omega)$. Within the statistical DMFT the full self-energy  $\Sigma_{ij}(\omega)$ is approximated by a self-energy diagonal in the lattice indices, i.e. $\Sigma_{ij}(\omega)=\delta_{ij} \Sigma_{i}(\omega)$. Within this approximation for interacting systems, Eq.~(\ref{eq_M1}) is modified to
\begin{equation}
G_{ii} (\omega) = \frac{1}{\omega + \mu -\epsilon_i - \Sigma_i(\omega) - \Gamma_i(\omega) + i \eta} \,. \label{eq_M4}
\end{equation} 
This approximation of the self-energy becomes exact in infinite dimensions, as was shown by Metzner and Vollhardt\cite{Metzner89}, and was used as a starting point for developing the DMFT.\cite{Vollhardt93,Georges96}

Explicitly, within the statistical DMFT Hubbard model (\ref{eq_sys1}) is mapped onto an ensemble of Anderson single impurity models (as schematically shown in Fig. \ref{figMethod}). We now repeat the previously described LD algorithm with the additional description on how to determine the self-energy of the interacting system.

In a fully interacting problem, different frequencies $\omega$ do not decouple in the self-consistency relations. This is in contrast to the non-interacting case, where the self-consistency equations are solved for each frequency separately. In the interacting case, we are therefore restricted to ensembles typically of the order $10^3$ samples. In order to reduce computation time, we also need to use a fast method (a so-called impurity solver) for determining the self-energy $\Sigma(\omega)$. Here, we use the iterative perturbation theory (IPT), \cite{MullerHartmann89,Potthof97,Kajueter97} which properly reproduces the non-interacting and atomic limits and was shown to qualitatively describe the Mott-Hubbard metal-insulator transition at a critical interaction $U$.\cite{Zhang93} Within the IPT the self-energy is calculated in second order in $U$ in the non-renormalized perturbation expansion. 

The original formulation of the IPT was restricted to the half-filled case. Later the method was extended to densities away from half-filling; this is commonly referred to as MPT.\cite{Kajueter97,Potthof97} The self-energy within MPT is given by\cite{Potthof97}
\begin{equation}
\Sigma(\omega) = U n + \frac{a \Sigma^{(2)}(\omega)}{1 - b \Sigma^{(2)}(\omega)} \,, \label{eq_M6}
\end{equation}
where 
\begin{equation}
a = \frac{n(1-n)}{n^{(0)}(1-n^{(0)})}
\end{equation}
and 
\begin{equation}
b =  \frac{B-B^{(0)}-\mu+\tilde{\mu}+ U(1-2n)}{U^2 n^{(0)}(1-n^{(0)})} \,,
\end{equation}
are additional coefficients in the interpolative formula. Here, $n^{(0)}$ denotes the filling obtained by using the Hartree-Fock solution
\begin{equation}
\rho^{(0)} (\omega) = -\frac{1}{\pi} \IM \big( \frac{1}{\omega+\tilde{\mu}-\epsilon-U n-\Gamma(\omega)+ i\eta}\big)  
\end{equation}
The parameter $\tilde{\mu}$ and the higher order correlation function $B^{(0)}$ and $B$ have to be fixed in such a way that the correct first three moments of the spectral density are guaranteed. In perturbation theory the second-order contribution to the self-energy is given by 
\begin{eqnarray}
\Sigma^{(2)}(\omega) &=&  \frac{U^2}{i} \int\limits_0^{\infty} dt \exp(i \omega t) \big( \tilde{\rho}_-(t) \tilde{\rho}_+(t) \tilde{\rho}_+(t) \nonumber\\
&&  + \tilde{\rho}_+(-t) \tilde{\rho}_-(-t) \tilde{\rho}_-(-t) \big)\, , \label{eq_M7}
\end{eqnarray}
where the Laplace transformed density of states is 
\begin{equation}
\tilde{\rho}_{\pm}(t) = \int\limits_0^{\infty} d\omega \exp(- i \omega t)  \rho^{(0)}(\pm\omega) \label{eq_M8} \,.
\end{equation}

According to Potthoff \textit{et al.}\cite{Potthof97} there are three approaches for fixing  $\tilde{\mu}$. The first is to require $\mu=\tilde{\mu}$. Second, one imposes the Friedel sum rule to ensure the low energy Fermi liquid behavior as done by Kajueter and Kotliar.\cite{Kajueter97} The last possibility requires that $n^{(0)} = n$. All three possibilities do not affect the validity of the MPT in the weakly interacting limit as all methods guarantee that $\tilde{\mu}\rightarrow \mu_{U=0}$ as $U \rightarrow0$.\cite{Potthof97} Furthermore, all three approaches have been compared carefully and checked against exact diagonalization (ED) calculations.\cite{Potthof97} In conclusion, the second and the third approaches show very good agreement whereas the first one differs considerably from ED results. In this work we choose the third possibility. 

The higher order correlation function $B^{(0)}$ is expressed as 
\begin{eqnarray}
B^{(0)} &=& \epsilon + \frac{1-2 n^{(0)}}{\pi n^{(0)}(1-n^{(0)})}  \IM\int\limits_{-\infty}^{0} d\omega \Gamma(\omega+i \eta)\nonumber \\ &&  \times G^{(0)}(\omega+i \eta) \,.
\end{eqnarray}
The correlation function $B$ is given by
\begin{eqnarray}
B &=& \epsilon-\frac{1}{\pi n(1-n)} \IM\int\limits_{-\infty}^{0} d\omega \Gamma(\omega+i \eta) \nonumber \\
&&  \times \big(\frac{2}{U} \Sigma(\omega+i \eta)-1 \big) G (\omega+i \eta) \,,
\end{eqnarray}
which can be solved self-consistently.

Studying correlated and disordered lattice fermions within statistical DMFT, we found that it is useful to compare the resulting spectral functions to those determined within the CPA.\cite{Soven67, Elliot74, Vlaming92} 
Within CPA the hybridization $\Gamma(\omega)$ is given by
\begin{equation}
\Gamma(\omega) = t^2 K [ x G_{\epsilon_i=-\Delta /2}(\omega) +  (1-x) G_{\epsilon_i=+\Delta /2}(\omega) ]\,.
\end{equation}
Note however, that the CPA is not able to describe Anderson localization.\cite{Lloyd,Th,wegner} 

\begin{figure}[tb]
\includegraphics[width=0.48\textwidth]{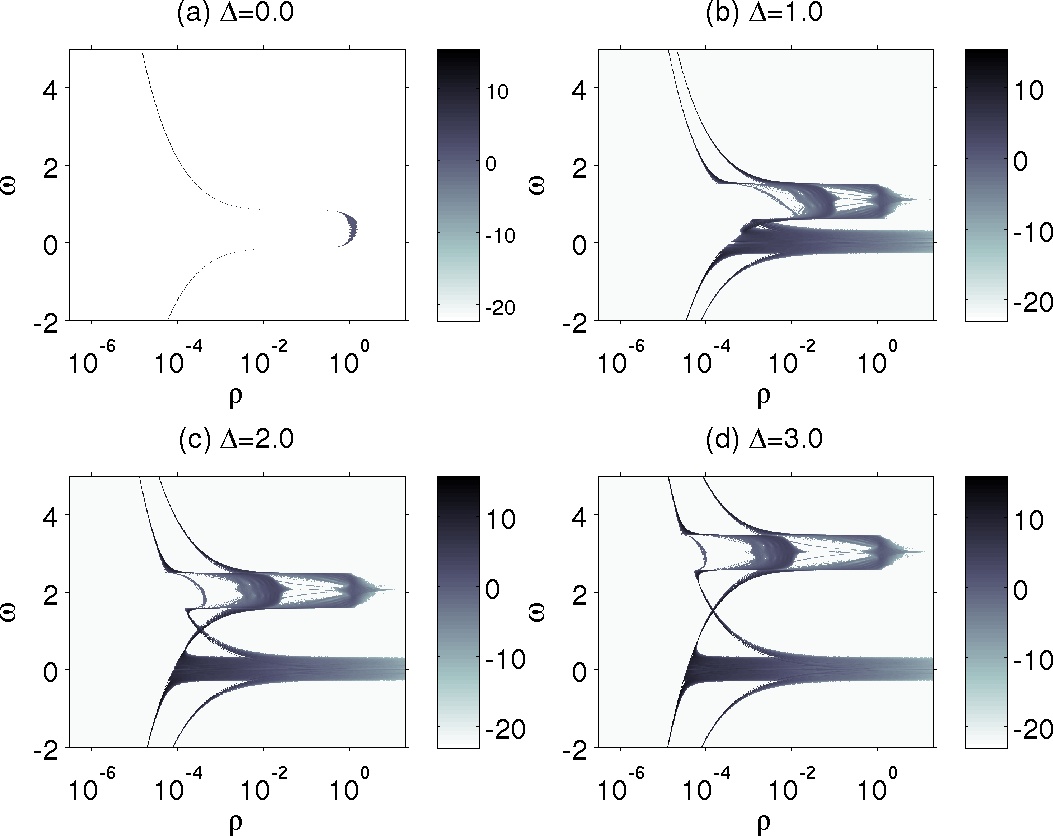}  
\caption{(Color online) Natural logarithm of the PDFs of the non-interacting system  $p[\rho]$ plotted color coded for fixed broadening $\eta=10^{-3}$ and  several disorder parameters $\Delta$: (a) $\Delta=0.0$,  (b) $\Delta=1.0$, (c) $\Delta=2.0$, (d) $\Delta=3.0$. Parameters are $K=6$, $\nu=0.2$,  $x=0.2$.}
\label{figPdelta}
\end{figure}

\section{Results}\label{results}

In this section we present our results obtained by statistical DMFT concerning Anderson and Mott-Hubbard transitions in correlated fermionic systems with binary-alloy disorder at zero temperature. In particular, we investigate how the predictions from Ref.~\onlinecite{Byczuk04} are extended when the Anderson localization is present. 

\begin{figure}[tb]
\includegraphics[width=0.48\textwidth]{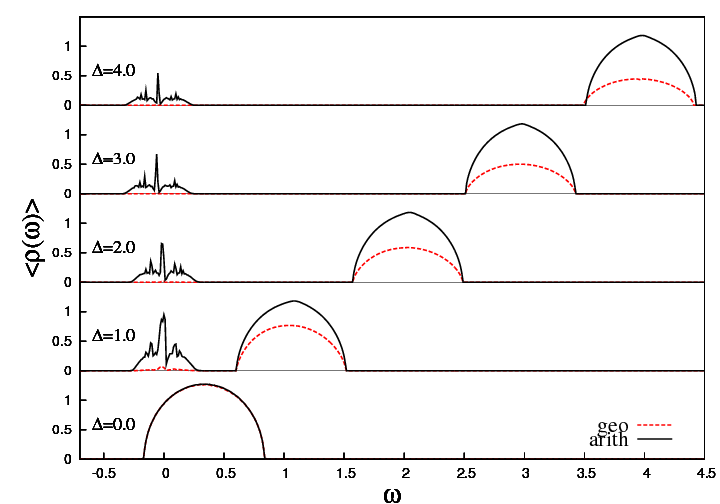}  
\caption{(Color online) Geometrically (dashed red line) and arithmetically averaged (solid black line) spectral functions of the non-interacting system for fixed broadening $\eta=10^{-3}$ and several disorder parameters ($\Delta=0.0, 1.0, 2.0, 3.0, 4.0$). Parameters are $K=6$, $\nu=0.2$, $x=0.2$.}
\label{figA}
\end{figure}
In the following we set the impurity concentration $x$ and the total particle density $\nu$ to be equal, i.e. $x=\nu$, by adjusting the chemical potential during the iterative solution of DMFT equations. This choice enables us to study Mott-Hubbard metal-insulator transition at non-integer particle densities. For practical calculations, we choose the impurity concentration and the particle density equal to $x=\nu=0.2$. Furthermore, we set the coordination number $K=6$ shortly above the classical percolation threshold $x_p=1/K$,\cite{Reich78} i.e., extended states can exist within both upper and lower alloy bands when they are split due to disorder.

\subsection{Detecting Anderson transition in the non-interacting case}

\begin{figure}[b]
\includegraphics[width=0.48\textwidth]{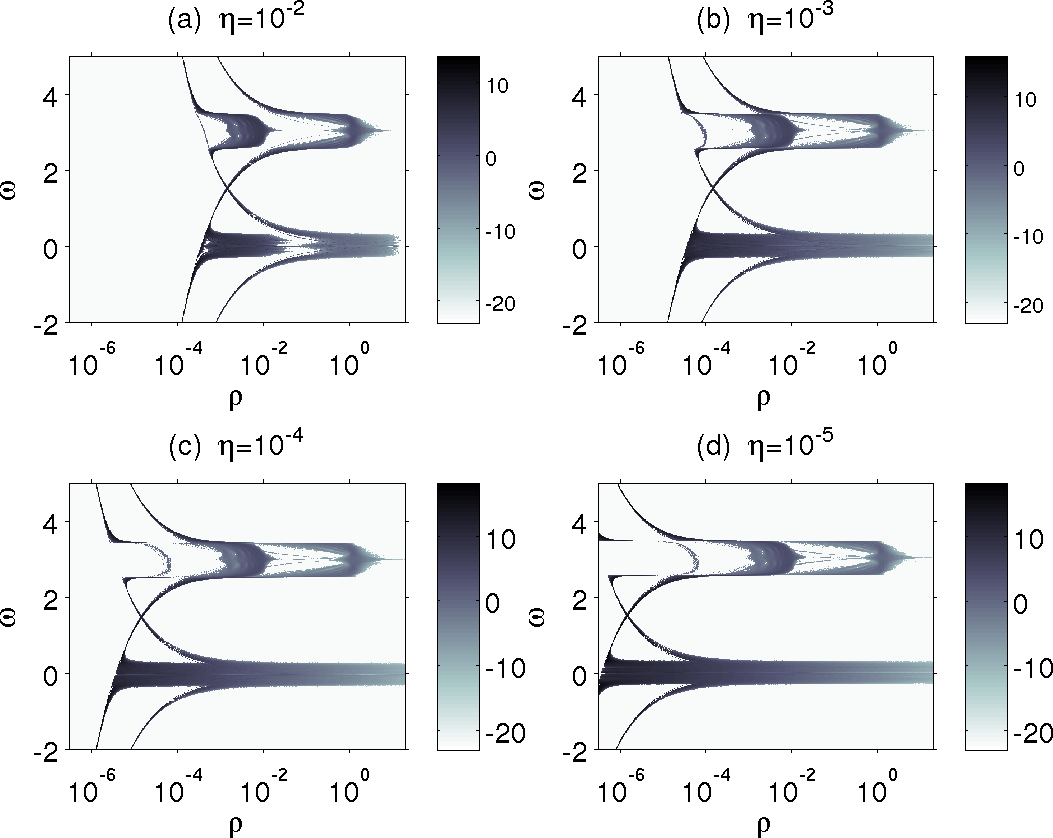}  
\caption{(Color online) Natural logarithm of the PDFs $p[\rho]$ of the non-interacting system plotted color coded for fixed disorder parameter $\Delta=3.0$ and several broadenings $\eta$: (a) $\eta=10^{-2}$,  (b) $\eta=10^{-3}$, (c) $\eta=10^{-4}$, (d) $\eta=10^{-5}$. Parameters are $K=6$, $\nu=0.2$, $x=0.2$.}
\label{figPeta}
\end{figure}

We first discuss how to detect localization effects and how to distinguish between extended and localized states in the non-interacting limit. Figure \ref{figPdelta} shows the PDFs on a logarithmic scale for different increasing values of the disorder parameter $\Delta$. A band splitting into an upper and a lower alloy band occurs with increasing disorder. This can also be seen plotting the arithmetically and geometrically averaged density of states as shown in Fig.~\ref{figA}. Looking at Fig.~\ref{figA}, we observe a vanishing geometrically averaged LDOS in the minority band. This corresponds to disappearing extended states and is used within TMT-DMFT to identify the Anderson transition. As mentioned before, in this paper we use a more powerful and general approach to detect Anderson localization. 

\begin{figure}[tb]
\includegraphics[width=0.48\textwidth]{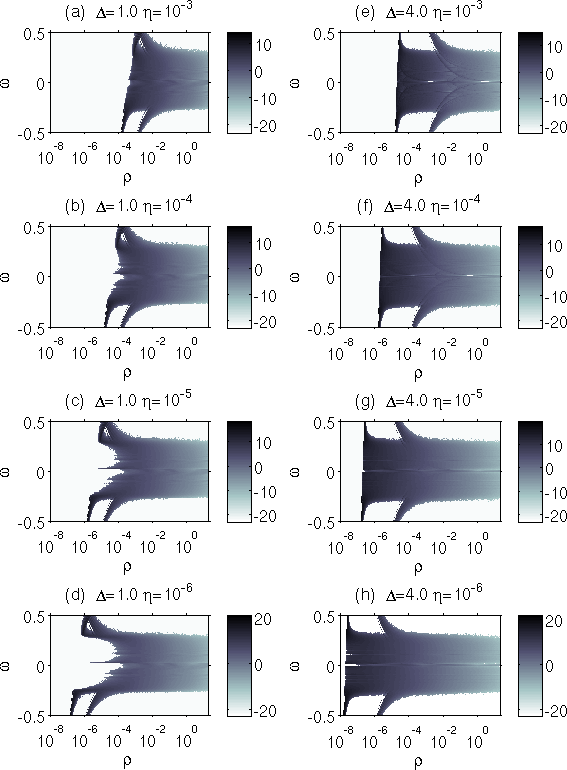}  
\caption{(Color online) Comparison of color coded natural logarithm of PDFs $p[\rho(\omega)]$ of the minority band of the non-interacting system for two different disorder parameters $\Delta=1.0$ (plots on the left side (a)-(d)), $\Delta=4.0$  (plots on the right side (e)-(h)) and for several broadenings $\eta$: (a,e) $\eta=10^{-3}$, (b,f) $\eta=10^{-4}$, (c,g) $\eta=10^{-5}$, (d,h) $\eta=10^{-6}$. Parameters are $K=6$, $\nu=0.2$, $x=0.2$.}
\label{figPminority}
\end{figure}

Extended states are characterized by a branch cut on the real axis of the local Green's function, whereas localized states are characterized by a dense distribution of poles in the thermodynamic limit.\cite{Economou73} This fact can be used to detect if states are localized or extended by investigating the behavior of the PDFs of the LDOS $p[\rho_i (\omega)]$ shown in Fig.~\ref{figPdelta} when the broadening $\eta$ tends to $0$.\cite{Alvermann05} Namely, the PDF of the LDOS for extended states saturates at a finite value for $\eta \rightarrow 0$, whereas the PDF of the LDOS for localized states decreases to zero for $\eta \rightarrow 0$. As an example, Fig.~\ref{figPeta} shows the behavior of the PDF when decreasing the broadening from $\eta=10^{-2}$ to $\eta=10^{-5}$ for a selected value $\Delta=3.0$. A change is seen for states in the lower alloy band, whereas the PDFs of the upper alloy band remain almost unchanged in this regime of $\eta$. The probability distributions of the LDOS of the lower alloy band are presented in detail for $\Delta=1.0$ and $\Delta=4.0$ in Fig.~\ref{figPminority}. It is clearly visible that the  PDFs for small $\Delta=1.0$,  corresponding to the lower alloy band, become $\eta$-independent for $\eta\rightarrow 0$. On the contrary, at large $\Delta =4.0$ the PDFs strongly depend on $\eta$.

\begin{figure}[tb]
\includegraphics[width=0.48\textwidth]{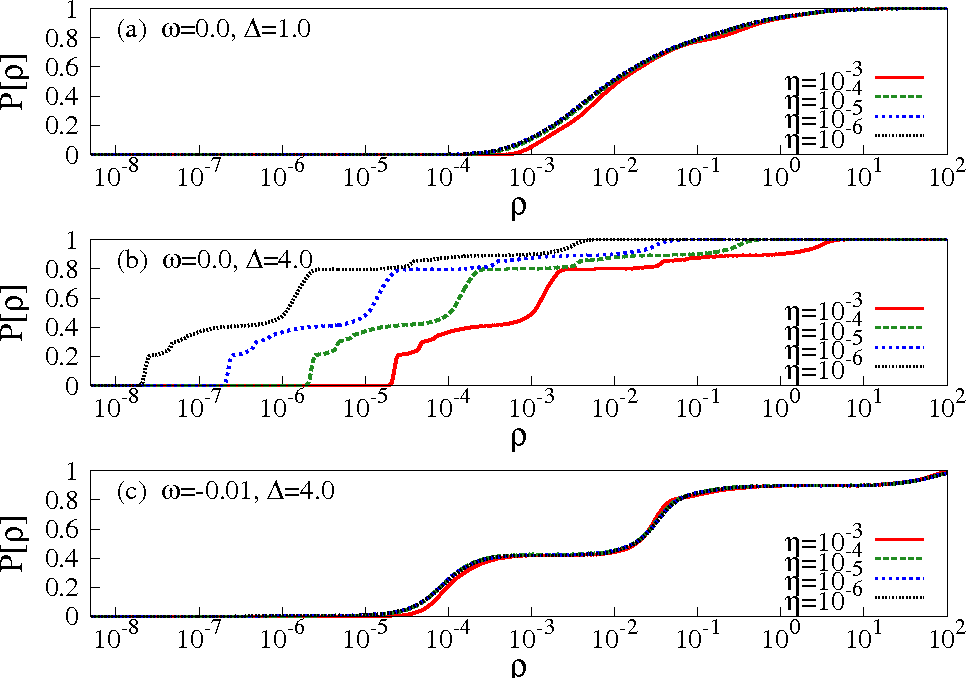}  
\caption{(Color online) Behavior of cumulative PDFs $P[\rho(\omega)]$ of the non-interacting system with decreasing broadening $\eta$ (a) for an extended state at $\Delta=1.0$ and $\omega=0.0$, (b) for an Anderson localized state at $\Delta=4.0$ and $\omega=0.0$  and (c) for an ``anomalous'' localized state at $\Delta=4.0$ and $\omega=-0.01$. Parameters are $K=6$, $\nu=0.2$, $x=0.2$.}
\label{figPsum_2D}
\end{figure}

In addition to Anderson localization effects we also observe that the spectrum is fragmented (cf. Figs. \ref{figA} and \ref{figPminority}), due to the presence of states with different physical properties. These states differ in the behavior of the PDF of the LDOS for  $\eta\rightarrow 0$ (cf. Fig. \ref{figPminority}) and are identified either as cluster resonances\cite{AlvermannAlloy05, Kirkpatrick72} or as ``anomalous`` localized states.\cite{Schubert05} The resonance states are similar bound eigenstates but with a finite life-time. They appear because of special geometrical configurations of the impurity atoms. The ''anomalous`` localized states are in fact extended states over the whole lattice but they are insulating and do not contribute to the dc conductivity.\cite{Schubert05} On a bipartite lattice these states have small wave function amplitudes on one sublattice and large amplitudes on the other sublattice. The typical $\eta$ behaviors of the PDFs for given frequencies are shown in Fig.~\ref{figPsum_2D}. Panel~(a) shows the behavior of an extended state, panel~(b) presents the behavior of an Anderson localized state, and panel~(c) shows the behavior of an ''anomalous'' localized state with its typical bimodal structure.\cite{Schubert05}

\subsection{Anderson and Mott transitions in the interacting case}

In the interacting limit we restrict our investigation of the $\eta$-dependence to the lower limit $\eta=10^{-5}$, as we use small ensembles due to computational limitations. We also note that the MPT requires a small finite broadening in any case.

We first comment on the defining properties of the different phases arising.
The paramagnetic metal is characterized by a non-vanishing arithmetically averaged LDOS at the Fermi level $\langle \rho (\omega) \rangle_{\mbox{\tiny arith}}$. The paramagnetic metal is gapless and hence compressible. Since it is also different from the Anderson-localized phase, the geometrically averaged LDOS at the Fermi level is finite as well. The Mott insulator possesses an excitation gap which is of the order of the interaction strength, and therefore this phase is characterized by a vanishing arithmetically averaged LDOS at the Fermi level $\langle \rho(\omega=0) \rangle_{\mbox{\tiny{arith}}}$.

In the presence of both interaction and disorder an Anderson insulator with localized one-particle wave functions is not well-defined anymore due to  many-body effects. Therefore, we refer to an Anderson-Mott insulator phase if the PDF of the LDOS tends to zero at the Fermi edge $\omega=0$ when $\eta\rightarrow 0$.

We also need to distinguish between an Anderson-Mott insulator and a band insulator. The band insulator is characterized by $\langle \rho(\omega=0)\rangle_{\mbox{\tiny{arith}}}=0$, but in contrast to the Mott insulator the excitation gap is determined by the energy distance between the upper edge of the occupied band and the upper alloy band, which in this case is proportional to $\Delta$.

\begin{figure}[tb]
\includegraphics[width=0.47\textwidth]{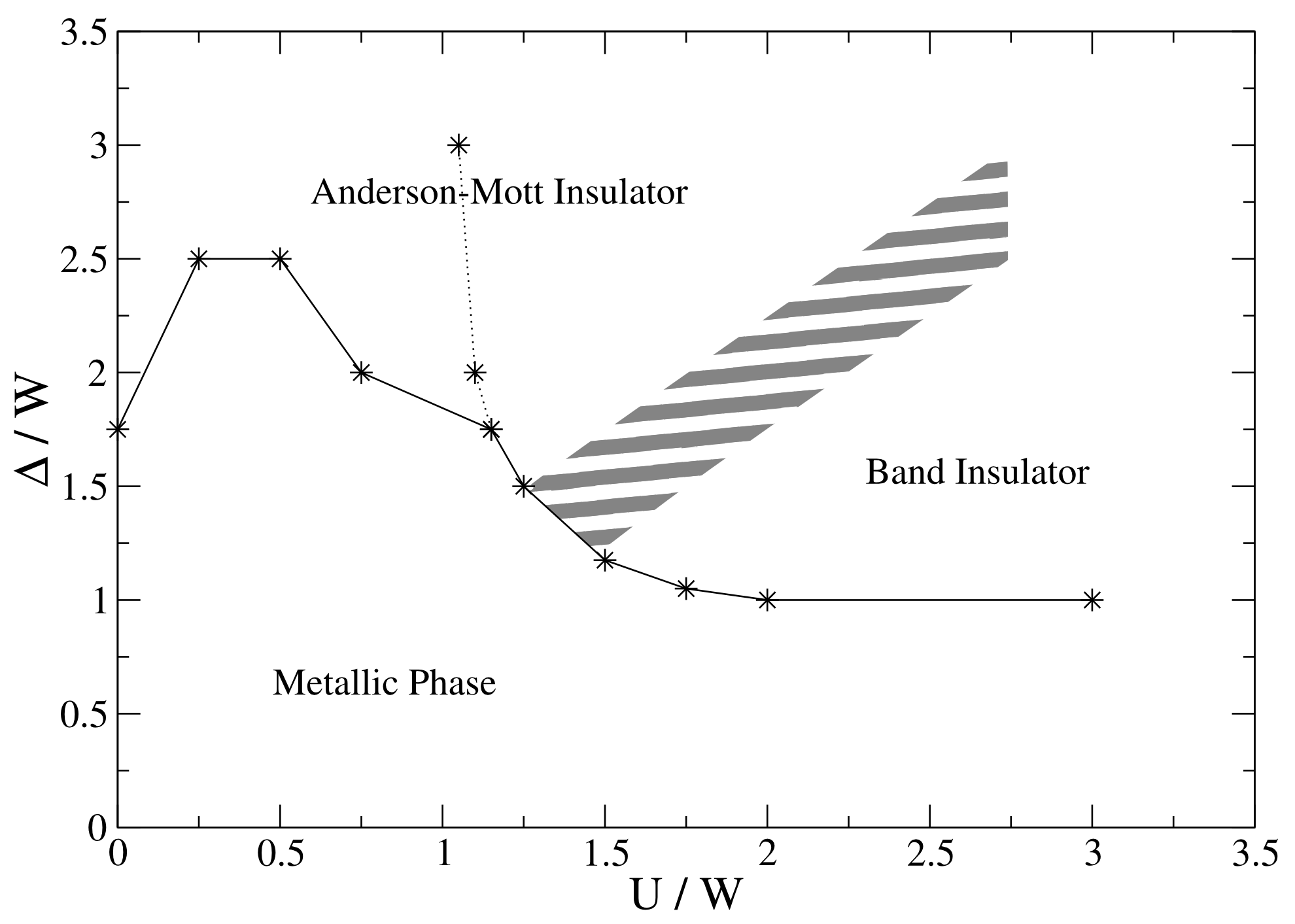}  
\caption{Phase diagram for the interacting and disordered system in $\Delta-U$-plane, showing of Anderson-Mott insulator, paramagnetic metal and band insulator. The solid line corresponds to the transition between insulating phases and the metal, the dotted line corresponds to a vanishing arithmetic average of the LDOS at the Fermi level and the dashed region denotes the crossover between Anderson-Mott insulator and band insulator. Parameters are $K=6$, $\nu=0.2$, $x=0.2$.}
\label{figPD}
\end{figure}

\begin{figure}[tb]
\includegraphics[width=0.48\textwidth]{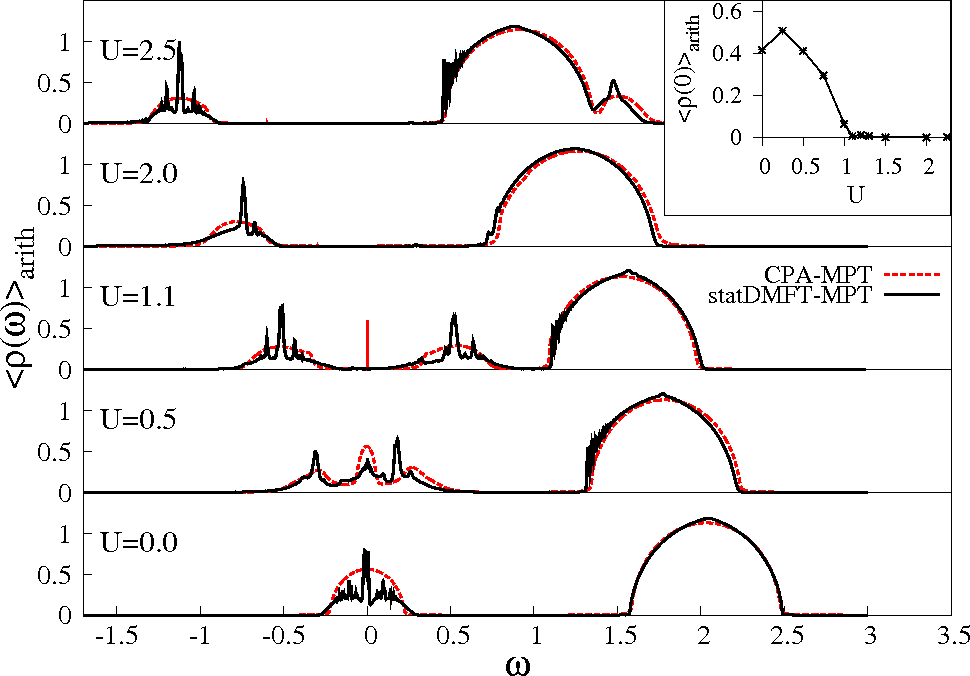}  
\caption{(Color online) Arithmetic average of the LDOS for increasing interaction strength $U$ ($U=0.0, 0.5, 1.1, 2.0, 2.5$) at fixed disorder parameter $\Delta=2.0$. The results from statistical DMFT calculations (solid black line) are compared to CPA results (dashed red line). The inset shows the arithmetic mean of the LDOS at the Fermi level with increasing interaction strength. Parameters are $K=6$, $\nu=0.2$, $x=0.2$.}
\label{figMIT_D2}
\end{figure}

The phase diagram presented in Fig.~\ref{figPD} is the main result of our paper. 
We find a metallic phase which turns into a Mott insulator at small $\Delta$ due to alloy band splitting and the mechanism described earlier in Refs.~\onlinecite{Byczuk03} and \onlinecite{Byczuk04}. However, as the current results prove, this type of Mott-Hubbard metal-insulator transition at non-integer particle densities is also possible, if Anderson localization effects are taken into account. In the limit of large disorder parameter $\Delta$ the metallic phase as well as the Mott-Hubbard transition is terminated by Anderson localization. The states in the upper part of the phase diagram in  Fig.~\ref{figPD} are localized due to strong disorder. 

\begin{figure}[b]
\includegraphics[width=0.48\textwidth]{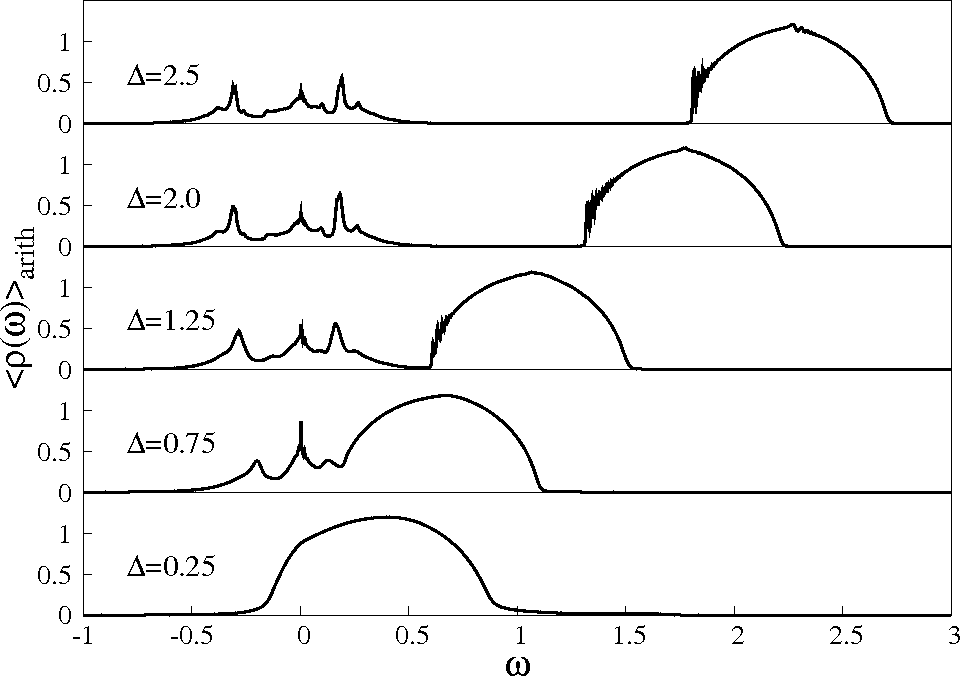}  
\caption{Arithmetic average of the LDOS with increasing disorder parameter $\Delta$ ($\Delta=0.25, 0.75, 1.25, 2.0, 2.5$) for fixed interaction strength $U=0.5$ and broadening $\eta=10^{-3}$. Parameters are $K=6$, $\nu=0.2$, $x=0.2$.}
\label{fig_inc_delta}
\end{figure}

Spectra corresponding to the Mott-Hubbard transition are displayed in Fig.~\ref{figMIT_D2}, where the arithmetically averaged LDOS obtained within statistical DMFT is compared to that obtained within a CPA type treatment of disorder. With increasing interaction $U$ at fixed $\Delta$ three peaks emerge because of the Mott-Hubbard and band splitting transitions. Moreover, we observe additional spikes in the LDOS similar to those observed for the non-interacting system.\cite{Alvermann05} These spikes are not reproduced by a CPA treatment of disorder and we conclude that they are due to local interference effects on clusters of impurity atoms.  
In the inset of Fig.~\ref{figMIT_D2}, the arithmetic average of the LDOS at the Fermi level is presented as a function of $U$. The Mott-Hubbard transition appears to take place at $U=1.1$. However, this is not a true transition point as it corresponds to the regime within the Anderson-Mott insulator where all states are already localized, cf.  Fig.~\ref{figPD}. We also see in Fig.~\ref{figMIT_D2} that by further increasing the interaction to $U=2.0$ the upper alloy band and the upper Hubbard band are merging. This corresponds to the crossover regime between alloy Anderson-Mott insulator and alloy-charge band insulator indicated by the dashed area in the phase diagram in Fig.~\ref{figPD}.\cite{Byczuk04} 
An additional effect is observed in  Fig.~\ref{figMIT_D2}, namely, with increasing $U$ the position of the upper alloy band is shifted with respect to zero on the energy scale, cf. Ref.~\onlinecite{Lombardo06}. This shift of the upper alloy band resembles a situation seen in the exactly solvable atomic limit.\cite{tobe}

For comparison we present the LDOS at a selected $U$ value for different disorder parameters $\Delta$ in Fig.~\ref{fig_inc_delta}. As expected, we observe a band splitting with increasing $\Delta$ and the formation of an energy gap proportional to $\Delta$ between the lower and the upper alloy bands. Note that additional peaks appear in the lower band when the disorder parameter $\Delta$ is increased. These peaks do not occur in a CPA treatment of disorder.

\begin{figure}[tb]
\includegraphics[width=0.48\textwidth]{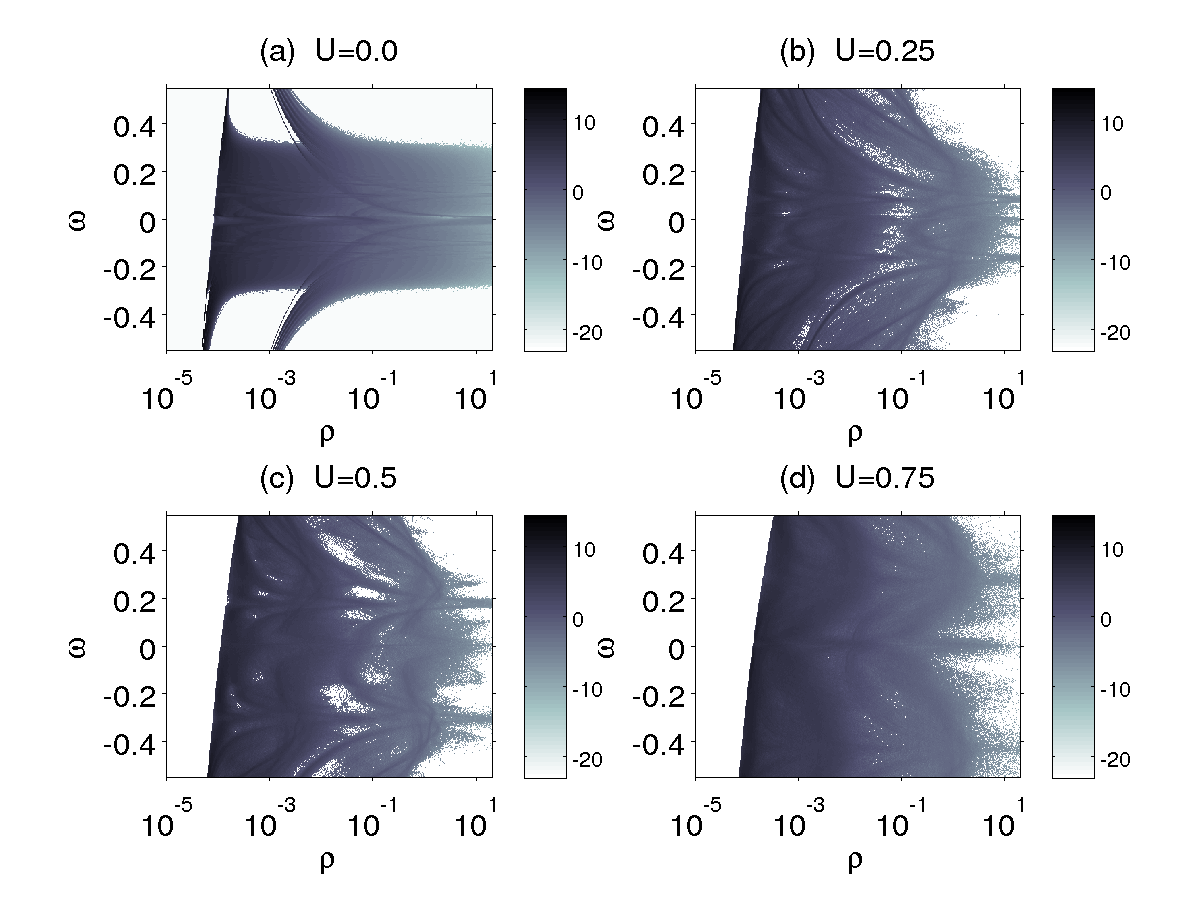}  
\caption{(Color online) Comparison of the color coded natural logarithm of PDFs $p[\rho(\omega)]$ for disorder parameter $\Delta=2.0$ and several interaction strengths $U$: (a) $U=0.0$,  (b) $U=0.25$, (c) $U=0.5$, (d) $U=0.75$. Parameters are $K=6$, $\nu=0.2$, $x=0.2$.}
\label{figMIT_D2_prob}
\end{figure}

Finally, in Fig.~\ref{figMIT_D2_prob} we show the evolution of the PDFs across the Mott-Hubbard transitions. The onset of a three peak structure is seen as well as sharp resonances in the LDOS being broadened and washed out by increasing the interaction strength $U$.

\section{Measuring the density of states in solids and ultracold Fermions}\label{connection_Exp}

In angle-resolved photoemission spectroscopy (ARPES) the photocurrent is given by\cite{Hopkinson80}
\begin{eqnarray}
I_{\mbox{\tiny ph}}(\omega) &=& -\frac{1}{\pi} \IM \int\int d \mathbf{r}d\mathbf{r}'\Psi_2(\mathbf{r},\omega+\Omega) \Phi(\mathbf{r}) \nonumber\\ &&\quad\quad \times G(\mathbf{r},\mathbf{r}',\omega)\Phi^*(\mathbf{r}')\Psi_2^*(\mathbf{r'},\omega+\Omega) ,\label{Iph}
\end{eqnarray}
with $G$ as the single-particle Green's function of energy $\omega$, the LEED state $\Psi_2$, the photon frequency $\Omega$, and the electron-photon interaction $\Phi$. If the investigated system is a disordered alloy, the ensemble average of the photocurrent has to be calculated.\cite{Durham81} In Ref.~\onlinecite{Durham81} it was shown that the ensemble averaged photocurrent is essentially given by the Bloch spectral function which is defined as\cite{Kudrnovsky86}
\begin{equation}
A(\omega,\mathbf{k}) = -\frac{1}{\pi} \IM \langle \mathbf{k}| \mbox{Tr} \langle G(\omega) \rangle | \mathbf{k} \rangle \,,
\end{equation}
where $\langle ... \rangle$ denotes the ensemble average. Here, the momentum $\mathbf{k}$ is a good quantum number as the ensemble averaging restores translational invariance.  Hence, in ARPES applied to alloys the Bloch spectral function is measured and therefore the ensemble averaged single-particle Green's function, which allows a comparison of the experimentally measured density of states to theoretical calculations (cf. Ref.~\onlinecite{Bansil83} and references therein).

Regarding experiments with cold atoms, a very promising probing technique, the momentum-resolved radio frequency spectroscopy,\cite{Stewart08} has been developed.
In the experiment\cite{Stewart08} a radio frequency field of frequency $\Omega_{\mbox{\tiny rf}}$ was applied to a two-component mixture of fermionic $^{40}$K atoms in hyperfine states $|1\rangle$ and $|2\rangle$ in order to excite the atoms of hyperfine state $|2\rangle$ to hyperfine state $|3\rangle$. The trap is then turned off and by counting the number of atoms in state $|3\rangle$, $N_3$, the dispersion $\epsilon_{\mathbf{k}}$ is obtained.\cite{Stewart08} The rf current, defined by $I=\langle\dot{N_3}\rangle$ is given by\cite{He05}
\begin{equation}
I(\mathbf{k}, \delta\nu) = \frac{|T_{\mathbf{k}}|^2}{2\pi} \rho(\mathbf{k},\omega) f(\omega)|_{\omega=\xi_{\mathbf{k}}-\delta\nu} \label{Irf}
\end{equation}
for homogeneous systems. $T_{\mathbf{k}}$ denotes the transition matrix, $\delta\nu$ is the rf detuning, and $\xi_{\mathbf{k}}$ is equal to $k^2/2m-\mu$, assuming that hyperfine state $|3\rangle$ is not occupied. In comparison, for homogeneous systems the photocurrent of Eq.(\ref{Iph}) reduces to\cite{arpes} 
\begin{equation}
I_{\mbox{\tiny ph}}(\mathbf{k},\omega) = M(\mathbf{k},\Omega) \rho(\mathbf{k},\omega) f(\omega) \,, 
\end{equation}
which, compared to Eq.(\ref{Irf}), reveals the analogy between ARPES and momentum-resolved rf spectroscopy for homogeneous systems. In inhomogeneous systems, such as trapped and/or disordered systems, final state effects have to be taken into account, which can be described by the density functional theory with the local density approximation.\cite{Chen09} In conclusion, the arithmetically averaged LDOS calculated in this work can in principle be compared to spectra resulting from momentum-resolved rf spectroscopy applied to fermions in an optical lattice. In order to realize such a comparison, the analog of Eq.~(\ref{Iph}) needs to be calculated for the RF current and the ensemble average has to be carried out accounting for final state effects.\cite{tobe}

\section{Summary}\label{summary}

We have investigated the binary alloy disordered Hubbard model within statistical DMFT, using MPT as an impurity solver. This method treats disorder and interaction on equal footing and in a non-perturbative way. The scheme reduces to the local distribution approach for non-interacting systems and to a standard  DMFT-MPT scheme in the pure case. Applying the statistical DMFT, we were able to compute the full probability distribution function of the local density of states, and therefore, localization effects have been studied in a more rigorous way and in more detail than in a typical medium theory combined with DMFT. As a result, the paramagnetic ground state phase diagram was obtained. It consists of a disordered metallic phase, an Anderson-Mott insulator, and a band insulator. For non-integer particle density $n=x$ a Mott-Hubbard metal-insulator transition is obtained even when Anderson localization effects are taken into account.

In future work, the method will be generalized to finite temperatures. It can also be extended to the hypercubic lattice. From the methodical point of view, it would be desirable to use impurity solvers that are superior or complementary to MPT.

\section*{Acknowledgments}
We acknowledge many useful discussions with D. Vollhardt and U. Bissbort. This work was supported by the Deutsche Forschungsgemeinschaft (DFG) via Forschergruppe No. FOR 801. Computations were performed at the Center for Scientific Computing of the Goethe University Frankfurt. K.B. also acknowledges support by the Sonderforschungsbereich No. 484 of the DFG and Grant No. N202026 32/0705 of the Polish Ministry of Science and Education.


\begin{thebibliography}{99}
\bibitem{comment_stat} In order to minimize the statistical fluctations we artifically increase the size of our ensemble after having reached the self-consistency. This is done by constructing the histogram on basis of typically $50-100$ successional update iterations.
\end{thebibliography}
\end{document}